\documentclass[grl]{agu2001}
\usepackage{lineno}

\pdfoutput=1
\usepackage{graphicx}
\slugcomment{Submitted to Geophysical Research Letters, \today}

\authorrunninghead{BALE ET AL.}
\titlerunninghead{NIF Electric potential at shocks}

\authoraddr{Stuart D. Bale,
Space Sciences Laboratory,
University of California,
Berkeley, CA  94720-7450, USA.
(bale@ssl.berkeley.edu)}

\pdfoutput=1

\begin{document}

\title{Direct measurement of the cross-shock electric potential at
low plasma $\beta$, quasi-perpendicular bow shocks}

\author{S. D. Bale\altaffilmark{1}, F. S. Mozer\altaffilmark{1}, and V. V. Krasnoselskikh\altaffilmark{2}}

 \altaffiltext{1}
 {Physics Department and Space Sciences Laboratory, University of California, Berkeley.}
 \altaffiltext{2}
 {Laboratoire de Physique et Chimie de l'Environnement, Orl\'eans, France.}

\begin{abstract}
We use the Cluster EFW experiment to measure the cross-shock electric field
at ten low $\beta$, quasi-perpendicular supercritical bow shock crossings on March 31,
2001.  The electric field data are Lorentz-tranformed to a Normal Incidence
frame (NIF), in which the incoming solar wind velocity is aligned with the shock normal.  In a 
boundary normal coordinate system, the cross-shock (normal) electric field is
integrated to obtain the cross shock potential.  Using this technique, we measure the
cross-shock potential at each of the four Cluster satellites and using an electric field
profile averaged between the four satellites.  Typical values are in the range 500-2500 
volts.  The cross-shock potential measurements are compared with the ion kinetic energy
change across the shock.  The cross-shock potential is measured to be
from 23 to 236\% of the ion energy change, with large variations between the four Cluster spacecraft at
the same shock.  These results indicate that solar wind flow through
the shock is likely to be variable in time and space and resulting structure of the
shock is therefore nonstationary.

\end{abstract}

\begin{article}

%
%

\section{Introduction}

At collisionless shocks, a large 'cross-shock' electric potential arises to 
oppose the incoming plasma flow.  This potential, and its corresponding
electric field, have the sense to repel incoming ions (which comprise the bulk of
the kinetic energy and momentum) and to reflect some ions, providing additional dissipation and also plays a role
in the redistribution of the upstream flow and magnetic field to the downstream state;
the physics of this energy partitioning is not yet fully understood. 

In the Ohm's law sense, the cross-shock electric field
arises from a combination of the Hall current, electron pressure gradients, and drag due to
the small population of gyrating ions at the shock front.  There are few, reliable published measurements
of the cross-shock potential, largely because it requires a DC (double-probe) electric field
instrument and these have been deployed primarily on low-apogee magnetospheric missions.
Formisano [1982] estimated a voltage drop of 140 and 240V at two shocks.  The double-probe
instrument on ISEE was used to measure electric fields of up to 100 mV/m at the shock [{\em Wygant et al.}, 1987] and
a Normal Incidence Frame (NIF) electric potential of 420V [{\em Scudder et al.}, 1986].  {\em Eastwood et al.} [2007]
used Cluster measurements to calculate a potential drop of 260V, which corresponded to the ion kinetic energy
change across the shock.  Very large electric fields (600 mV/m), including parallel electric fields of 100 mV/m
have been measured by the Polar spacecraft at a high Mach shock [{\em Bale and Mozer}, 2007], however the single-spacecraft Polar mission
does not allow for good estimates of shock frame transformations.  Furthermore, the bow shock is known to have large
amplitude electric field structure from electron inertial scales [{\em Walker et al.}, 2004] down to Debye scales
[{\em Bale et al.}, 1998].
Several authors have used a Liouville mapping of thermal
electrons across the bow shock to estimate the deHoffman-Teller frame shock potential [viz. {\em Schwartz et al.}, 1988], which
can be related back to the NIF potential by a frame transformation.  {\em Gedalin} [1997] calculated the expected shock
potential (assuming pressure balance) and found that it should peak at an
Alfven Mach number of around 2; this is where the magnetic compression begins to saturate, while the shock thickness
continues to grow [{\em Bale et al.}, 2003] with Mach numbers, leading to a small potential and more magnetic reflection at
high Mach numbers.

The four Cluster spacecraft can be used to calculate a shock reference frame, which is important
to transform correctly electric field data.  In this letter, we measure directly the cross-shock electric field
and compute the NIF electric potential.  We find that the potential varies significantly between the
different spacecraft at the same shock, indicating rapid temporal and/or spatial variations.  This seems to be 
consistent with expectations of shock 'reformation' [e.g. {\em Krasnoselskikh et al.}, 2002; {\em Hellinger et al.}, 2002
{\em Matsukiyo and Scholer}, 2006] and indicates that plasma flow through the shock
and particle heating and energization are likely to be bursty.

\section{Cluster data}
The four Cluster spacecraft fly together in a controlled tetrahedron
orbit with apogee near 19 $R_e$ and inter-spacecraft
separations that vary from a few hundred to several thousand kilometers.  Each winter,
apogee passes through the dayside of the magnetosphere and Cluster
crosses the bow shock (at least twice during each 57 hour orbit).
On March 31, 2001, Cluster encountered the bow shock 11 times as a CME/magnetic cloud passed over the earth.  The large,
steady magnetic field of the cloud gives a low upstream plasma $\beta$ and hence,
very planar bow shocks, which are ideal for this study.  The alpha particle density
was especially high during this interval, with an average value of $n_\alpha/n_p \approx 9\%$,
as is often the case within magnetic clouds.  The alpha density was used in computing ion masses
 below; however, it is not clear how the enhanced alpha density will effect the
measured electric potentials.

The EFW experiment [Gustafsson et al., 1997] measures probe-to-spacecraft voltage on four 8 cm spherical voltage probes, each extended on wire booms 44 m from the spacecraft body, in the spin plane.  Since only two components of the electric field are
measured, we assume that $\vec{E} \cdot \vec{B}$ = 0 (ideal MHD) in order to determine the three-component electric field and correct this where required later.  Since the 
$B_z$ (GSE) component is large here (in a CME), this makes for a good reconstruction of the missing ($E_z$) component
of the electric field.  The electric field data are calibrated locally near each shock by forcing agreement between the component
of the ion velocity perpendicular to $\vec{B}$ and $(\vec{E} \times \vec{B})/B^2$; this correction is of order 1 mV/m in the $X$ GSE
direction and minimizes any offsets due to varying plasma
or photoelectron/secondary electron conditions.  The sum of the four probe voltages gives an estimate of the spacecraft floating
potential which is related functionally to the ambient plasma density.  We fit the spacecraft potential
locally (near each shock) to the plasma density to produce a high time resolution 'density proxy' measurement.
We also use magnetic field data from the FGM experiment [Balogh et al., 1997], ion moments
from the CIS instrument [Reme et al., 1997], and electron temperatures from the PEACE instrument [Johnstone et al., 1997].  The fractional alpha particle density is estimated using
ACE data upstream and convected back to the shock crossing time; the alpha density is used to estimate
the solar wind mass density, where needed.

%
%

 \begin{figure}[h]
\centering
 \includegraphics[width=85mm,
	height=75mm, scale=1,clip=true, draft=false]{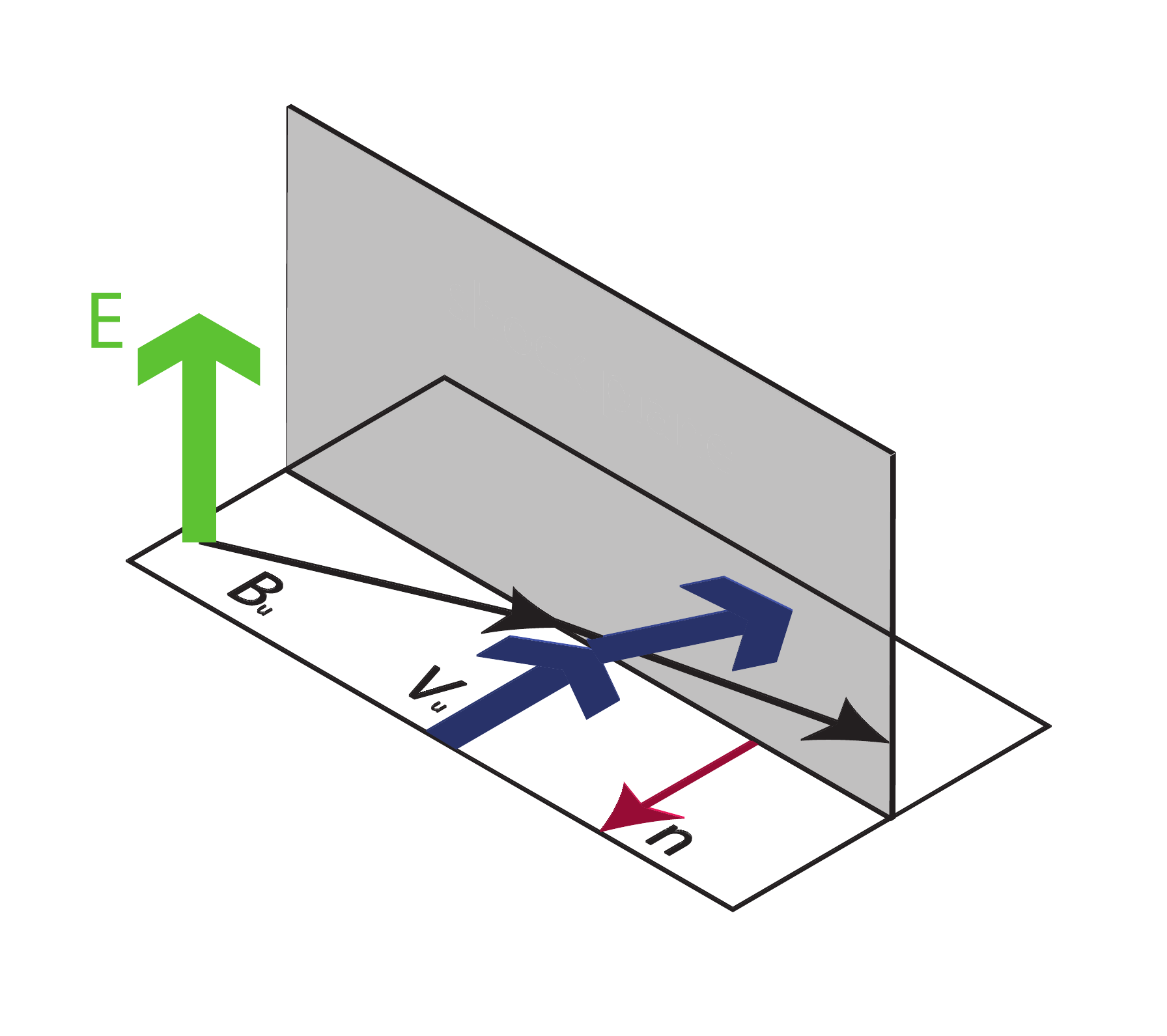}
\caption{The shock coordinate system is defined with respect to the
normal and the coplanarity plane, which contains the magnetic field and
velocity vectors.  In the normal incidence frame, the solar
wind velocity is directed along the normal (red arrow).  The convection electric field
$\vec{E} = - \vec{v} \times \vec{B}$ (green arrow) is normal to the coplanarity plane
and is approximately constant through the shock layer.}
\end{figure}

\section{Measurements at bow shock crossings}

Of the 11 bow shock crossings observed by Cluster on March 31, 2001, ten
(10) of them were chosen for this study.  These shocks were used by {\em Maksimovic et al.}
[2003] to study global shock motion.

\subsection{Shock normals}
We compute the shock normal $\hat{n}$ by comparing the shock arrival time at the
four Cluster satellites $\tau_i$ and inverting the matrix equation $\mathbf{r} \cdot \hat{n} = 
v_{sh} \vec{\tau}$, where $\mathbf{r}$ is a matrix of relative spacecraft positions and $v_{sh}$
is the shock speed in the spacecraft frame [viz {Bale et al.}, 2003].  We do this separately using
both spacecraft potential and magnetic field magnitude data and find normals that agree to within
a few degrees and speeds that agree to within a few km/s, typically.  Minimum variance normals are
also in good agreement.  A single spacecraft crosses
the shock in 2-10 seconds typically and the shock transit time between different spacecraft is from
7 to 37 seconds.

%
%

\begin{figure}[h]
\centering
 \includegraphics[width=82mm,
	height=105mm, scale=1,clip=true, draft=false]{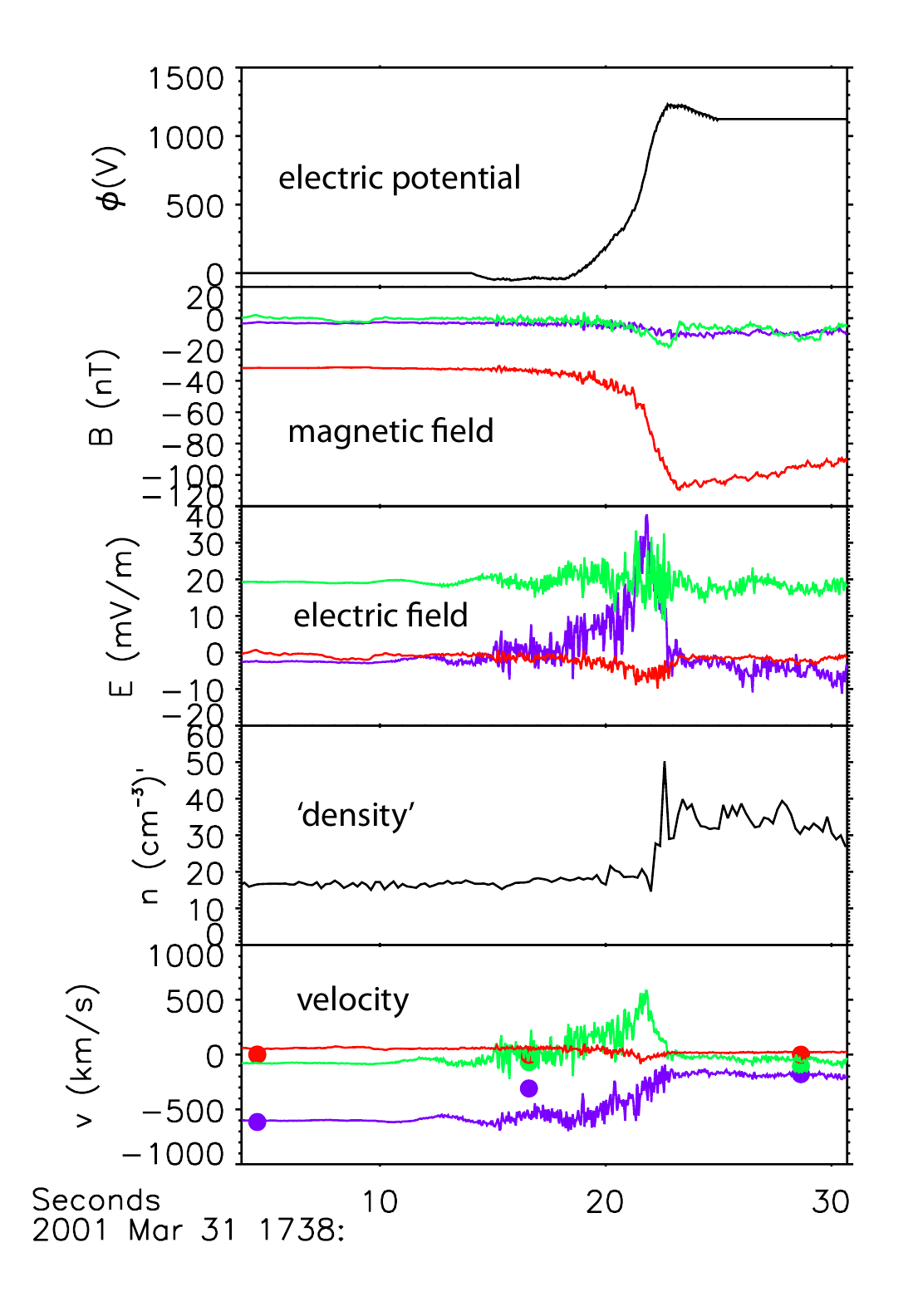}
\caption{A typical shock crossing on March 31, 2001.  The electric and magnetic fields
and perpendicular velocity are in the coordinate system described in Figure 1; the purple line is the $\hat{x}$ component,
green is $\hat{y}$, and red is the $\hat{z}$ component, large dots in the bottom panel show the perpendicular
component of the ion velocity measured by CIS.  The electric potential (top panel) is computed by
integrating the normal (purple) component of the electric field (middle panel).  The $B_y$ (out-of-coplanarity) magnetic
 field can be seen in the second panel (green trace).}
\end{figure}


\subsection{Lorentz frame and coordinate transformations}
The spacecraft-frame (measured) electric field can differ from that of the shock frame by
several mV/m.  Therefore, we Lorentz transform the measured field into the shock frame using
the measured shock velocity $\vec{v}_{sh}=v_{sh} \hat{n}$ to generate the shock-frame electric field $\vec{E}_{sh} = 
\vec{E} + \vec{v}_{sh} \times \vec{B}$.  Then we compute the NIF velocity $\vec{v_{NIF}} = \hat{n} \times
(\vec{v}_u \times \hat{n})$, where $\vec{v}_u$ is the solar wind velocity in the shock frame, and Lorentz
transform the electric field $\vec{E}_{NIF} = \vec{E}_{sh} + \vec{v}_{NIF} \times \vec{B}$ and the incoming
flow velocity $v_{sw} = v_{u}-v_{NIF}$ to the NI frame.
Note that to this point, we have assumed that $\vec{E} \cdot \vec{B} = 0$ (which is a Lorentz invariant).

Finally, the shock frame electric and magnetic fields and velocity are rotated into a coordinate system (Figure 1) which is defined with the shock normal
as the $\hat{x}$ direction and the maximum variance of the magnetic field vector as the $\hat{z}$ direction; $\hat{x}$ and
$\hat{z}$ define the coplanarity plane.  In this coordinate system, the
cross-shock electric field is in the $\hat{x}$ direction, the convection electric field $-\vec{v}_{sw} \times \vec{B}$ is in the $\hat{y}$ direction,
and the magnetic field shearing direction is $\hat{z}$.  Figure 2 shows data in this frame and coordinate system at one of our shock crossings.

\subsection{The cross-shock electric potential}
The shock speed calculated as described in Section 3.1 is used to generate a spatial shock
profile $dx = v_{sh} dt$ and the cross-shock ($\hat{x}$ component) electric field can be integrated directly to obtain
the cross-shock potential.  To compensate for the assumption of $\vec{E} \cdot \hat{B} = 0$,
we now assume that the true cross-shock electric field lies purely in the $\hat{x}$ (normal) direction,  and that
we are measuring only the projection of it perpendicular to $\hat{B}$; therefore the cross-shock field can be
written as $\vec{E}_s = E_{s} ~\hat{x} = {\underline{(E_s \sin\Theta_{bn}})} \hat{\perp} - (E_s \cos\Theta_{bn})\hat{\parallel}$,
where $\hat{\parallel}$ and $\hat{\perp}$ are unit vectors parallel and perpendicular to the magnetic field (and hence
in the coplanarity plane).  The perpendicular component (underlined above) is the measured $E_{\perp,x}$ and therefore we can recover the cross-shock field amplitude as $E_s = E_{\perp,x}/\sin\Theta_{bn}$ and this is the field that we integrate to obtain
the NIF potential; this represent a correction of from $0.1\%$ for $\Theta_{bn} \approx 87^\circ$ to 
$21\%$ for $\Theta_{bn} \approx 56^\circ$ (see Table 1).  It is interesting to note that the cross-shock electric field is comprised of both a large-scale DC field and shorter
wavelength, spiky structures of comparable amplitude (viz [{\em Walker et al.}, 2004]).  These structures are included in the integral of cross-shock
potential, however, the resulting potential profile is relatively smooth and monotonic (until the magnetic overshoot, which
is mimicked in the electric potential profile).

At each shock, the electric potential $\phi_{SCi}$ is computed, as described above, on ${\it each}$ of the
four Cluster spacecraft ($i=1,4$) and an 'average' potential $\phi_A$ is computed by first aligning (in time) the data
from the four spacecraft, computing an average cross-shock electric field profile, and then integrating
it, so that $\phi_A$ is like a potential of the volume-averaged field.  Table 1 lists the 10 shocks, their
macroscopic parameters, the Alfven Mach number $M_A$, electron and proton plasma beta, the shock tangent
angle $\Theta_{bn}$, and the ion energy $E = 1/2 m v^2_u$ and its change across the shock $\Delta E = 1/2 ~m (v_u^2-v_d^2)$, along
with the measured electric potentials.

%
%

\begin{table*}[b]
\def\ph{\phantom{age }}
\def\pph{\phantom{th}}
\newbox\dothis
\setbox\dothis=\vbox to0pt{\vskip-1pt\hsize=11.5pc\centering
Error\vss} \caption{Normal Incidence Frame (NIF) shock parameters and measured cross-shock electric potentials on March 31, 2001. }
\begin{tabular*}{\textwidth}{@{\extracolsep{\fill}}ccccrrrcrrcccc}
\hline
\cline{3-6}\cline{7-10}\cr
\noalign{\vskip-2pt}
& & &&&\multicolumn{2}{c}{ }&\multicolumn{5}{c}{cross-shock potential (volts)}\cr 
\noalign{\vskip 3pt}
\cline{8-{12}}\cr
\noalign{\vskip-16pt}
&\multicolumn{1}{c}{ }&\multicolumn{1}{c}{ }&\multicolumn{1}{c}{ \hbox to-5pt{}}&&&\multicolumn{1}{c}{ }&\multicolumn{1}{c}{ }&\cr
Shock Time&\multicolumn{1}{c}{$M_A$} &\multicolumn{1}{c}{$\beta_e$}&\multicolumn{1}{c}{$\beta_p$}&
\multicolumn{1}{c}{$\Theta_{bn}$\hbox to-12pt{}}&E (eV)\tablenotemark{a}&$\Delta$E (eV)\tablenotemark{b}&%
\multicolumn{1}{c}{$\phi_{SC1}$} &\multicolumn{1}{c}{$\phi_{SC2}$}&
\multicolumn{1}{c}{$\phi_{SC3}$\hbox to -12pt{}}&$\phi_{SC4}$&${\phi_A}$\tablenotemark{c}&${{\phi_A}}/{E}$&${{\phi_A}}/{\Delta E}$\cr
\noalign{\vskip3pt} \hline \noalign{\vskip3pt} 
17:14:45&2.4& 0.03&0.02\ &83$^\circ$ & 2299 &1926\ph& 1973&1900\pph&2057&3245\ph & 2190 &0.95 & 1.14 \cr
17:18:50&2.9& 0.01& 0.01\ &85$^\circ$ & 2223 &2116\ph &1200&830\pph&1754&1232\ph& 1223 &0.55 & 0.58  \cr 
17:36:47&3.2& 0.06& 0.05\ &86$^\circ$& 2458 &2110\ph  & 251&535\pph&825&755\ph & 482 &0.20 & 0.23 \cr 
17:38:20&3.9& 0.05&0.06\ &86$^\circ$ & 2561 &2331\ph  & 840&1884\pph&923&1375\ph & 1239 &0.48 & 0.53 \cr
18:02:15&3.4& 0.03& 0.03\ &87$^\circ$ & 2818 &2467\ph  &893&714\pph&1126&971\ph & 909 &0.32& 0.37 \cr 
18:28:40&5.5& 0.10 & 0.10\ &84$^\circ$ & 2487 & 2266\ph  & 2373&1541\pph&1185&2348\ph & 1800 &0.72& 0.79 \cr 
18:48:20&2.5& 0.11&0.07\ &57$^\circ$ & 2581 & 2121\ph  & 1748&1127\pph&920&1146\ph& 1039 &0.40& 0.49\cr
19:00:41&3.7& 0.10 &0.09\ &64$^\circ$ & 2362 & 2053\ph  & 2968&2813\pph&3623&2992\ph& 2791 &1.18& 1.36\cr
19:46:37&2.6& 0.02&0.02\ &62$^\circ$ & 2261 & 1798\ph  & 1012&809\pph&726&980\ph& 799 &0.35& 0.44\cr
21:34:06&2.7& 0.03&0.03\ &56$^\circ$ & 1820 & 1688\ph  & 4303&4987\pph&3495&5402\ph& 3992 &2.19& 2.36\cr
\noalign{\vskip 3pt}
\hline
\end{tabular*}
\tablenotetext{a}{$E = 1/2 m v^2_u$ is the NIF upstream ion kinetic energy}
\tablenotetext{b}{$\Delta E = 1/2 m (v^2_u-v^2_d)$ is the NIF ion kinetic energy change from upstream to downstream}
\tablenotetext{c}{Note that ${\phi_A}$ is not the average of the four potentials, rather it is the
cross-shock potential of the {\em average} shock electric field profile.}
\end{table*}

\section{Discussion}

The last columns of Table 1 show $\phi_A/E$ and $\phi_A/\Delta E$, the average cross-shock potential normalized to
the upstream ion kinetic energy $E=1/2 m v^2_u$ and its change across the shock $\Delta E = 1/2 m (v_u^2-v_d^2)$.  This is a measure of the ability of the shock to oppose the directed plasma
flow; i.e. when the electric potential  $\phi_A$ approaches the upstream ion energy $E$ the shock should turn back the {\em entire} solar wind thermal ion
population.  Figure 3 shows $\phi/E$ with the black dots as $\phi_A$ and the error bars representing the maximum
and minimum $\phi_i$ at each shock and plotted against Alfven Mach number; red dots show the values of  $\phi_A/\Delta E$ The electric potential can vary by
nearly $100\%$ in some cases and in three of the ten cases, $\phi_A/E$ is greater than 1.

%
%

\begin{figure}[h]
\centering
 \includegraphics[width=80mm,
	height=75mm, scale=1,clip=true, draft=false]{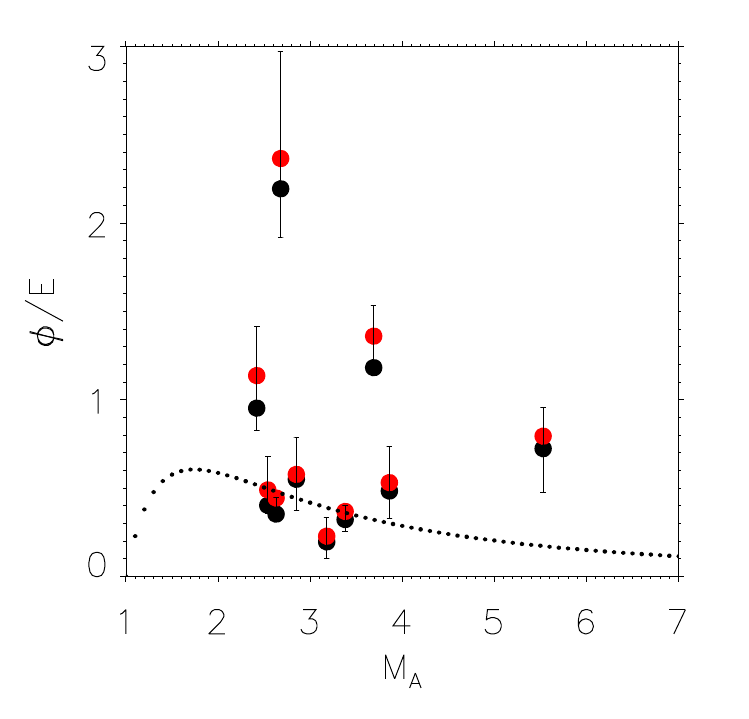}
\caption{The cross-shock potential normalized to the upstream NIF ion kinetic energy $E = \phi/(1/2 m v_u^2)$ plotted against Alfv\'en Mach
number.  The black dots are computed from the 'average' shock potentials $\phi_A$ and the 
error bars show the range of potential variation between the 4 Cluster satellites.  Red dots show the ratio of $\phi_A$
to the NIF ion kinetic energy change $\Delta E = 1/2 m (v_u^2-v_d^2)$.  The dotted line is the analytical relationship from {\em Gedalin} (1997).}
\end{figure}

{\em Gedalin} [1997] estimated the cross-shock potential in the NIF analytically
(assuming a monotonic shock profile and pressure balance) and found
that the cross-shock potential peaks at small Alfven Mach numbers ($M_A \approx$ 2), qualitatively in 
agreement with dHT potentials inferred from Liouville mapping of thermal electrons [{\em Schwartz et al.,}, 1988].
Our NIF potentials show a similar trend (Figure 3), albeit with poor statistics; the dHT and NIF cross-shock ($\hat{x}$) electric fields are 
related by a frame transformation $E_{dHT} = E_{NIF} + (E_y/B_x) B_y$.

Theoretical studies of shock reformation or nonstationarity predict that quasiperpendicular shock fronts should
be unstable in certain regimes of Mach number and plasma $\beta$ [e.g. {\em Krasnoselskikh et al}, 2002; {\em Hellinger et al.}, 2002; {\em Matsukiyo and Scholer}, 2006].
In particular, the constraint that the incoming solar wind speed be larger than the whistler phase speed and/or the ion thermal
speeds allows the development of a shock front instability that results in 'overturning' of the front with a characteristic
timescale of the ion gyroperiod and a spatial scale of the ion gyroradius.

Recently {\em Lobzin et al.} [2007], using Cluster observations, provided convincing evidence that
high Mach number quasiperpendicular shocks are nonstationary, moreover, a quasi-periodic shock
 front reformation takes place which modulates the reflected ion population.

In the reformation scenario, the shock electric potential will oscillate to large values on timescales of the ion gyroperiod $\tau_{ci}$.  
The time
between shock crossings here (from spacecraft to spacecraft) ranges from 7 to 30 seconds, while the ion cyclotron period is $\tau_{ci} \approx$
2 s; so it is plausible that these shocks are reforming rapidly and the four Cluster spacecraft each encounter the same shock at a different
phase of the reformation cycle resulting in large variations in the measured potential.  This strongly varying electric potential should
produce a modulated reflected ion flux, as observed by {\em Lobzin et al.} [2007].  During this interval, the Cluster spacecraft were separated by
400-900 km, while $\rho_i \approx$ 200 km, so that the spacecraft-to-spacecraft variations may be spatial, rather than temporal.

This is the first multi-spacecraft study of the cross-shock electric potential (to our knowledge) and it shows that
shock electric potentials vary largely on the timescale of the ion gyroperiod and/or spatial scales of ion gyroradii and
that the Normal Incidence frame potential is often observed to be greater than the ion kinetic energy change across
the shock.  These observations are consistent with the quasiperpendicular shock reformation scenario and suggest that
the transmission and reflection at the shock front are a bursty phenomena.  This shock reformation may be due to an inherent
instability (as described in references above) or due to solar wind driving, although there is no one-to-one signature of this
in the upstream data [{\em Maksimovic et al.}, 2003].

\begin{acknowledgments}
Cluster EFW data analysis at SSL is supported by NASA grant NNG05GL27G to
the University of California.
We acknowledge the Cluster CIS, PEACE, and FGM and the ACE EPAM teams for data.
\end{acknowledgments}

%
%

%
%

\end{article}

\end{document}